# GAS COMPRESSION MODERATES FLAME ACCELERATION IN DEFLAGRATION-TO-DETONATION TRANSITION

**Vitaly Bychkov***

Department of Physics, Umeå University, 901 87 Umeå, Sweden

**Damir Valiev, V'yacheslav Akkerman and Chung K. Law**

Department of Mechanical and Aerospace Engineering, Princeton University,
Princeton, NJ 08544-5263, USA

## Abstract

The effect of gas compression at the developed stages of flame acceleration in smooth-wall and obstructed channels is studied. We demonstrate analytically that gas compression moderates the acceleration rate and perform numerical simulations within the problem of flame transition to detonation. It is shown that flame acceleration undergoes three distinctive stages: 1) initial exponential acceleration in the incompressible regime, 2) moderation of the acceleration process due to gas compression, so that the exponential acceleration state goes over to a much slower one, 3) eventual saturation to a steady (or statistically-steady) high-speed deflagration velocity, which may be correlated with the Chapman-Jouguet deflagration speed. The possibility of deflagration-to-detonation transition is demonstrated.

*vitaliy.bychkov@physics.umu.se

**Introduction**

In spite of its extreme importance, for a long time the deflagration-to-detonation transition (DDT) remained one of the least understood processes of hydrodynamics and combustion science (Urtiev & Oppenheim, 1966; Oppenheim, 1985; Zeldovich *et al.* 1985; Landau & Lifshitz, 1989; Shepherd & Lee, 1992; Kerampran *et al.* 2000; Cooper *et al.* 2001; Law, 2006). Being crucial in the terrestrial conditions, the DDT event can occur in unconventional, unbounded astrophysical systems such as supernovae explosions (Oran, 2005; Mazalli *et al.*, 2007; Akkerman *et al.* 2011; Poludnenko *et al.* 2011; Gao & Law, 2011). Recently, pseudo-combustion phenomena of front acceleration and the DDT have been also obtained in advanced materials such as organic semiconductors and crystals of nanomagnets (Decelle *et al.*, 2009; Bychkov *et al.*, 2011; Modestov *et al.*, 2011 a,b). During the process of DDT, a usual slow flame (deflagration) accelerates spontaneously with velocity increase by three orders of magnitude until an explosion occurs and develops into a self-sustained detonation (Ott *et al.* 2003; Kagan & Sivashinsky 2003; Roy *et al.* 2004; Kuznetsov *et al.* 2005; Ciccarelli *et al.* 2005; Frolov *et al.* 2007; Oran & Gamezo, 2007; Gamezo *et al.* 2008; Ciccarelli & Dorofeev, 2008; Johansen & Ciccarelli, 2009; Dorofeev 2011; Finigan *et al.* 2011). The first qualitative explanation of the flame acceleration in tubes, involving the thermal expansion of the burning gas, non-slip at the tube walls and turbulence as the main components acceleration, has been suggested by Shelkin in 1940-ies (Shelkin, 1940). Namely, when a flame propagates from a closed tube end, the burning gas expands and pushes a flow of the fuel mixture, as shown schematically in Fig. 1. Due to of non-slip walls, the flow becomes strongly non-uniform, hence making the flame shape curved and thereby increasing the burning rate and driving the acceleration. Turbulence provides additional distortion of the flame front and compensates for thermal losses to the walls, and the acceleration of turbulent flames has been observed in numerous experiments (Roy *et al.* 2004; Kuznetsov *et al.* 2005; Frolov *et al.* 2007; Gamezo *et* *al.* 2008; Johansen & Ciccarelli, 2009). Despite a century of intensive research, turbulence in general and turbulent burning in particular belong to the most difficult problems of modern physics (Peters *et al.* 2000), and because of the complications related to turbulent burning, for a long time there was almost no progress in the quantitative theoretical understanding of the flame acceleration process.

A considerable step in the understanding the flame acceleration has recently been achieved within the approach of a laminar flow, based on the direct numerical simulations and the analytical theory supporting each other. The theory of laminar flame acceleration in smooth tubes has been developed and validated computationally by Bychkov *et al.* (2005; 2007), Akkerman *et al.* (2006). In tubes with obstacles Bychkov *et al.* (2008) have demonstrated that delayed burning between the obstacles creates a powerful jet-flow driving the acceleration which is much stronger than that according to the classical Shelkin scenario, see also Valiev *et al.* (2010). Thereby Bychkov *et al.* (2008) identified a new extremely fast mechanism of flame acceleration, independent of the Reynolds number, with turbulence playing only a supplementary role. Nevertheless, in both configurations of smooth-wall and obstructed tubes/channels, the theories of flame acceleration (Bychkov *et al.* 2005; 2007; 2008; Akkerman *et al.* 2006; Valiev *et al.* 2010) employed the limit of an incompressible flow, which holds with a good accuracy only at the beginning of the process. Indeed, a typical value of the unstretched laminar flame speed $U_f$ for hydrocarbon flames is about 40 cm/s, so the initial values of the Mach number related to flame propagation are quite small $Ma \equiv U_f / c_s \cong 10^{-3}$, where $c_s$ is the sound speed. Hence, the effects of gas compression may be neglected at the initial stages of the acceleration.

Recent experiments have confirmed the possibility of acceleration and DDT for ethylene-oxygen flames in micro-tubes with diameters about 1 mm (Wu *et al.* 2007; 2011) as well as for flames in micro-gaps (Wu & Kuo, 2011). Nevertheless, while the theories (Bychkov *et al.* 2005; 2007; 2008; Akkerman *et al.* 2006; Valiev *et al.* 2010) predicted the exponential acceleration of laminar flames in micro-scale tubes at the initial stage, Wu *et al.* (2007) demonstrated a number of specific effects beyond the scope of the incompressible flow models of the these theories, such as the saturation of the flame velocity to a steady value. The steady regime can be interpreted as the Chapman-Jouguet (CJ) deflagration (Landau & Lifshitz, 1989; Chue *et al.* 1993), being subsonic with respect to the fuel mixture just ahead of the flame front and supersonic in the reference frame of the tube walls. Similar saturation of the flame propagation speed to a supersonic value with respect to an observer has been detected experimentally in channels with obstacles; this regime is often called "fast flames" (Kuznetsov et al. 2005; Ciccarelli & Dorofeev, 2008). In order to elucidate these effects, we have to account for gas compression in both configurations of smooth-wall and obstructed tubes/channels.

In this paper, we report the recent results on the influence of gas compression on flame acceleration at the developed stages in both geometries. It is demonstrated analytically that gas compression moderates the acceleration rate. We also report the results of direct numerical simulations within the problem of flame transition to detonation. It is shown that the flame acceleration undergoes three distinctive stages: 1) initial exponential acceleration in the incompressible regime, 2) moderation of the acceleration process due to gas compression; consequently, the exponential acceleration regime goes over to a much slower one, 3) eventual saturation to a steady high-speed deflagration velocity. The saturation deflagration velocity may be correlated with the CJ deflagration speed. The possibility of DDT is demonstrated.

**The role of gas compression in moderating flame acceleration in smooth tubes**

Figure 1 illustrates schematically the Shelkin mechanism of flame acceleration in tubes with smooth walls, as well as the moderating role of gas compression in the process. At the incompressible stage, burning involves the drop of the gas density by a factor $\Theta \equiv \rho_f / \rho_b$, which is typically rather large. For the burning rate $U_w$ (roughly, the unstretched laminar flame speed $U_f$ multiplied by a scaled increase in the flame surface area due to curvature), the flame front produces the extra volume of the gas $(\Theta - 1)U_w$ per unit time. At the initial incompressible stage of flame acceleration, the burnt gas is mostly at rest due to the boundary conditions at the closed tube end, so that the extra volume results in a flow of the fuel mixture only. Accounting for gas compression, we obtain a counter-flow in the burnt matter in addition to the main flow in the fuel mixture. Initially, the role of the counter-flow is as small as $Ma \ll 1$, but it grows as the flame accelerates. Quantitative theory of the flame acceleration in a smooth-wall, two-dimensional channel of half-width $R$ has been developed by Bychkov *et al.* (2010a) considering the influence of gas compression as the Taylor series for a small parameter $Ma \ll 1$. According to (Bychkov *et al.* 2010a), the velocity $\dot{Z}_f$, found for the average position $Z_f(t) = \langle z_f(x,t) \rangle$ of the elongated flame front $z = z_f(x,t)$, obeys the differential equation

$$\ddot{Z}_f = \frac{U_f}{R}\sigma \dot{Z}_f \left(1 - Ma\, B \frac{\dot{Z}_f}{U_f}\right), \tag{1}$$

with the scaled acceleration rate $\sigma$,

$$\sigma = \frac{(\mathrm{Re}-1)^2}{\mathrm{Re}} \left(\sqrt{1 + \frac{4\,\mathrm{Re}\,\Theta}{(\mathrm{Re}-1)^2}} - 1\right)^2, \tag{2}$$

characterizing the exponential flame acceleration, $\dot{Z}_f \propto U_f \exp(\sigma U_f t / R)$, at the initial incompressible stage, $\mathrm{Re} = U_f R/\nu$ playing the role of the Reynolds number related to planar flame propagation, and a numerical factor found $B$ given by

$$B = \frac{\Theta - 1}{\Theta}\left[(\gamma - 1)\frac{\Theta - 1}{\Theta} + 1\right]\frac{1 + S}{1 - (\Theta - 1)S}, \qquad S = \frac{\sqrt{\sigma \mathrm{Re}}}{(\sqrt{2\sigma \mathrm{Re}} - 1)(\sqrt{\sigma \mathrm{Re}} + \sqrt{2\sigma})}, \tag{3}$$

where $\gamma$ is the adiabatic exponent. According to Eq. (1), the flame accelerates in the exponential regime at the initial stage, as long as the front velocity is strongly subsonic, $\dot{Z}_f / c_s \ll 1$. As the flame velocity approaches the sound speed, the role of gas compression, posed by the term $\propto Ma \dot{Z}_f / U_f$, increases and thereby moderates the acceleration regime. Qualitatively, the solution to Eq. (1) describes the transition from the initial exponential regime of flame acceleration to almost linear acceleration and then to saturation of the flame velocity as

$$\dot{Z}_f / U_f = \frac{\Theta \exp(\sigma U_f t / R)}{1 + Ma B \Theta \exp(\sigma U_f t / R)}. \qquad (4)$$

However, quantitatively, Eqs. (1) – (4) hold only as long as the term $Ma B \Theta \exp(\sigma U_f t / R)$ is small and can be treated as a correction in Taylor series, $Ma B \Theta \exp(\sigma U_f t / R) \ll 1$. Figure 2 compares the analytical theory, Eqs. (1) – (4) to the numerical simulations (Bychkov *et al.* 2005), with good agreement between the theory and the simulations as long as the flame speed is relatively small.

The role of gas compression at the developed stages of flame acceleration has been demonstrated in (Valiev *et al.* 2010) using direct numerical simulations. Figure 3 presents the tip velocity of the reaction front versus time for $\text{Re} = 6.7$, $Ma = 10^{-3}$ and $\Theta = 8$; the plot demonstrates all elements of the DDT from the initial flame acceleration to the steady detonation. Focusing at the acceleration process, we observe several distinctive stages: 1) initially, the flame accelerates exponentially in an isobaric (incompressible) regime; 2) later, the acceleration regime moderates to an approximately linear velocity increase; 3) subsequently, the flame velocity saturates to the quasi-steady regime with supersonic velocities in the laboratory reference frame. The saturation velocity is comparable to the CJ deflagration speed. The effect of gas compression, both behind the flame front and ahead of the front, can be observed directly in Fig. 4, which presents the density and velocity profiles along the channel axis at various time instants. The compression of the burnt gas behind the flame front is relatively uniform in agreement with the theory (Bychkov et al. 2010a), while in the fresh fuel mixture we can see a non-uniform, adiabatic compression wave and a shock pushed by the flame front. When the flame tip reaches the distance about $Z_{tip} \cong 3 \cdot 10^3 R$ from the closed end of the channel, the density of the fuel mixture in the compression wave exceeds its initial value approximately $3 - 4$ times. Maximal possible gas compression that could be achieved in a shock wave is $(\gamma + 1)/(\gamma - 1)$ (Chue *et al.* 1993), which equals 6 in the present case. The velocity distribution in Fig. 4 shows also the region of the counter-flow (negative velocity) behind the flame front, which tends to moderate the acceleration.

We also demonstrate that the flame acceleration leads finally to the detonation triggering. The whole multi-dimensional picture of the final stage of the DDT is shown in Fig. 5, with the color representation for the temperature. Figure 5 (a) presents all elements of the flame dynamics at that stage, while Figs. 5 (b) and (c) illustrate some interesting features of the process in detail. In Fig. 5 (a) we squeeze the pictures in z-direction to make the whole flow structure visible (we remind that the channel width is 2*R*). The central part in the first snapshot shows the elongated flame front at the very beginning of the explosion. In addition, we can see the explosion starting along the walls because of viscous heating as explained in (Valiev et al. 2009). The process is more pronounced thereafter, when the tongues of the explosion burst along the walls at high speed, catch up with the flame tip (second snapshot) and then leave it far behind engulfing the flame (third snapshot). Interaction of the explosion and the flame produces a strong turbulent flow, which enhances burning. Figures 5 (b) and (c) indicate that turbulence develops as a result of hydrodynamic instabilities, presumably, the Kelvin-Helmholtz, Rayleigh-Taylor and Richtmyer-Meshkov instabilities. We can recognize the classical elements of these instabilities: small perturbations at the beginning in Fig. 5 (b), a vortex street, "cat-eye" vortices and the "mushroom"-shape of the leading part of the flame front in Fig. 5 (c). Configuration of the turbulent burning region resembles the characteristic shape of an accelerating turbulent flame observed in the experiments (Kuznetsov *et al.* 2005; Ciccarelli & Dorofeev, 2008) quite well. Experimental papers typically describe such a process as fast turbulent burning in a boundary layer, which pushes a strong shock, thereby reducing the reaction time in the fuel mixture and facilitating the explosion. Since the shock is almost planar, the explosion spreads from the channel walls to the axis and produces detonation considerably ahead of the turbulent flame brush. Again, we emphasize strong resemblance between the present simulations and the scenario of "explosion-within-explosion" revealed experimentally (Ciccarelli & Dorofeev, 2008). A large pocket of unburnt gas remains trapped behind the detonation front. The detonation is seen on the last snapshot of Fig. 5 (a).

**Moderation of flame acceleration in obstructed tubes/channels due to gas compression**

As demonstrated in (Bychkov *et al.* 2008; Valiev *et al.* 2010), the physical mechanism of flame acceleration in obstructed tubes/channels is qualitatively different from the classical Shelkin scenario (Bychkov *et al.* 2005; Akkerman *et al.* 2006). This new acceleration mechanism is extremely strong, providing flame acceleration that is independent of the Reynolds number, and as such may be quite important for technical applications. Specifically, fast flame propagation in the free central part of an obstructed channel creates pockets of fresh fuel mixture between the obstacles, as illustrated in Fig. 6. Gas expansion due to delayed burning in the pockets produces a powerful jet flow in the unobstructed part of the channel. The jet flow renders the flame tip to propagate even faster, which produces new pockets, generates a positive feedback between the flame and the flow, and leads to the flame acceleration. According to the theory (Bychkov *et al.* 2008), developed within the limit of incompressible flow, the flame tip propagation is described by the equation

$$\dot{Z}_{tip} = \frac{(\Theta-1)}{(1-\alpha)} \frac{U_f}{R} Z_{tip} + \Theta U_f, \qquad (5)$$

yielding the exponential acceleration, $\dot{Z}_{tip} \propto \exp(\sigma U_f t / R)$, with the scaled acceleration rate $(\Theta-1)/(1-\alpha)$ independent of the Reynolds number, where $\alpha$ denotes the blockage ratio.

Similar to smooth tubes, the flame acceleration occurs because of the extra gas volume produced in the burning process and indicated by the factor $(\Theta - 1)$ in Eq. (5). As long as gas compression is negligible (at the initial stage of flame acceleration), this extra volume results in the jet flow shown in Fig. 6. However, as the speed of the flame tip approaches the sound speed, the effects of gas compression become important thereby making the jet flow in Fig. 6 weaker and moderating the acceleration process. The quantitative theory of the moderation mechanism has been developed in (Bychkov *et al.* 2010b) using the Taylor series for $Ma \ll 1$. Accounting for small, but finite gas compression we can extend the theoretical results (Bychkov *et al.* 2008) to

$$\dot{Z}_{tip} = \sigma_1 Z_{tip} - Ma \frac{\sigma_0^2}{U_f}\left(\frac{1}{1-\alpha} + \gamma - 1\right)Z_{tip}^2 + \Theta\left(1 - Ma(\gamma-1)\frac{(\Theta-1)^2}{\Theta}\right)U_f, \qquad (6)$$

where

$$\sigma_1 = \sigma_0\left[1 - Ma\left(\frac{\Theta}{1-\alpha} + 2(\gamma-1)(\Theta-1)\right)\right], \qquad \sigma_0 = \frac{(\Theta-1)U_f}{(1-\alpha)R}. \qquad (7)$$

Similar to Eq. (1) for smooth-wall channels, the derived Eq. (6) describes the moderating influence of gas compression, incorporated both in linear and nonlinear terms in $Z_{tip}$, in obstructed channels. Figure 7 compares the analytical results obtained for the incompressible flow, Eq. (5), the weakly compressible flow, Eq. (6), and the numerical data of (Bychkov *et al.* 2008; Valiev *et al.* 2010). In all three cases, the theory developed for non-zero Mach numbers agrees well with the simulation results at the initial stage of flame acceleration, but deviates at later stages. The states of deviation approximately correspond to the same level of flow compressibility, $u_z/c_s \approx 0.1$, which is achieved faster for larger values of the blockage ratio, e.g. at $Z_{tip}/R \approx 4.4; 6.6; 8.8$ for $\alpha = 1/3; 1/2; 2/3$, respectively. Still, the flame accelerates extremely fast in obstructed channels, thereby making the validity of the formulation based on the Taylor series quite limited in time. Moreover, in line with previous experimental and numerical studies (Kuznetsov et al. 2005; Frolov et al. 2007; Ciccarelli & Dorofeev, 2008; Johansen & Ciccarelli, 2009; 2010), at the later stage of the flame acceleration, the flow becomes essentially non-laminar, and the increase in vorticity leads to the modification of the flame tip velocity.

In order to study the role of gas compression at the developed stages of flame acceleration in obstructed channels, direct numerical simulations of the Navier-Stokes combustion equations have been performed (Bychkov *et al.* 2008; Valiev *et al.* 2010). Characteristic temperature and velocity distribution at the initial stage of the process are shown in Fig. 8, where we easily recognize the main elements of the new acceleration mechanism, namely, fast spreading of the flame fronts in the central free part of the channel, delayed burning in the pockets and the strong jet flow. Figure 9 shows the position of the flame tip versus time, scaled according to Eq. (5), as predicted by the theoretical model of incompressible flow and found in numerical simulations for various initial values of the Mach number. As we can see, the limit of incompressible flow holds with good accuracy for $Ma = 10^{-3}$, while the deviations are noticeable already for $Ma = 5 \cdot 10^{-3}$, and they are even stronger for $Ma = 10^{-2}$, with the effective acceleration rate smaller by a factor of about 2 as compared to the predictions of Eq. (5). Still, all plots of Fig. 9 demonstrate almost exponential acceleration of the flame tip versus time, which corresponds to a relatively initial stage of flame acceleration. Modification of the acceleration regime occurs at the later stages of the process as presented in Fig. 10. The moderation of flame acceleration due to gas compression agrees with the concept that the flame propagation velocity cannot exceed the limiting value of the CJ deflagration speed, for which the downstream flow is sonic. We therefore expect saturation of the flame tip velocity to a certain steady value at the end of the acceleration process, but prior to an explosion. Indeed, Fig. 10 demonstrates such saturation, obtained computationally at the final stage of flame acceleration, for various blockage ratios.

Finally, we discuss how flame acceleration in obstructed channels may lead to DDT. It is well known that any flamefront propagating from a closed end pushes a flow in the fuel mixture with a weak shock/compression wave at the head of the flow. The flame acceleration renders the compression wave stronger, until it develops into a shock of considerable amplitude. Preheating of the fuel mixture by the shock is conventionally considered as one of the main elements of DDT both in obstructed and unobstructed tubes/channels (Shelkin, 1940; Roy *et al.* 2004; Ciccarelli & Dorofeev, 2008). The temperature behind the shock increases, and the reaction time in any compressed gas parcel decreases drastically. The decrease in the reaction time may result in explosion and DDT ahead of the flame front unless the parcel is burnt by the flame before active explosion is initiated. Thus, in general, we may expect two possible outcomes for the flame acceleration: 1) if the reaction time behind the shock is sufficiently short, then it drives the explosion and DDT; 2) the reaction time may be longer than the interval available for a gas parcel to travel between the shock and the flame. In the latter case, the explosion does not occur and the final state of flame acceleration is the CJ deflagration. It is noted that both CJ detonation and deflagration have been experimentally found in smooth tubes (Wu *et al.* 2007). In the numerical simulations for the geometry of obstructed channels, we also observed both possibilities of DDT and CJ deflagration for different reaction kinetics. Taking reaction of the first order with respect to density (designated by $n=1$ in Fig. 10), we have obtained a statistically steady CJ deflagration at the end of flame acceleration with no explosion or DDT. This result indicates that the decrease in the reaction time behind the shock is not sufficient, and the gas parcels are consumed by the flame front before spontaneous reaction develops into a powerful explosion. Thus, in order to observe DDT, we needed to take another reaction order, e.g. $n=2$, which is more sensitive to pressure and temperature increase in the shock, and obtained explosion triggering and DDT, see Fig. 10. Remarkably, in this case the reaction rate is so sensitive to pressure and temperature that the DDT occurs before the flame reaches the CJ deflagration state.

**Summary**

We have considered the role of gas compression in moderating flame acceleration in DDT in the configurations of smooth-wall and obstructed tubes/channels. At the beginning the flow is almost incompressible, and we obtain quasi-isobaric the flame acceleration in an exponential regime. However, by expanding the theory to the first-order terms in the Taylor series for Mach number, we find that gas compression modifies the exponential regime into a much slower one. The developed stages of flame acceleration with considerable gas compression have been studied using direct numerical simulations. The numerical modelling substantiates predictions of the analytical theory, and shows moderation of the acceleration regime and eventual saturation to steady (or statistically-steady) fast flame propagation, which can be associated with the CJ deflagration known from the classical theory (Chue *et al.* 1993). We also demonstrate the possibility of DDT both for smooth-wall and obstructed channels.

**Acknowledgements**

This work was mostly supported by the Swedish Research Council (VR) and the Swedish Kempe Foundation. The numerical simulation was performed at the High Performance Computer Center North (HPC2N), Umea, Sweden. The work at Princeton University was supported by the US Air Force Office of Scientific Research.

**FIGURES**

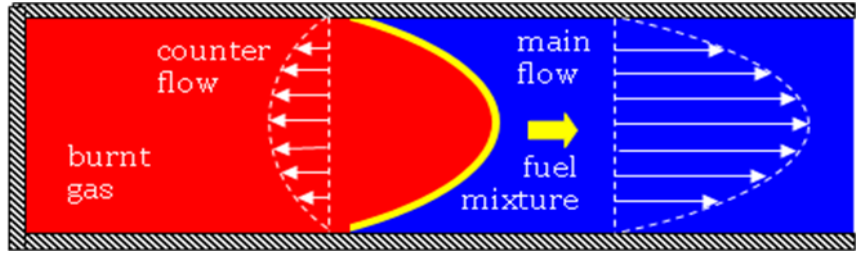

**Figure 1**. Schematic of the Shelkin mechanism of flame acceleration and the influence of gas compression

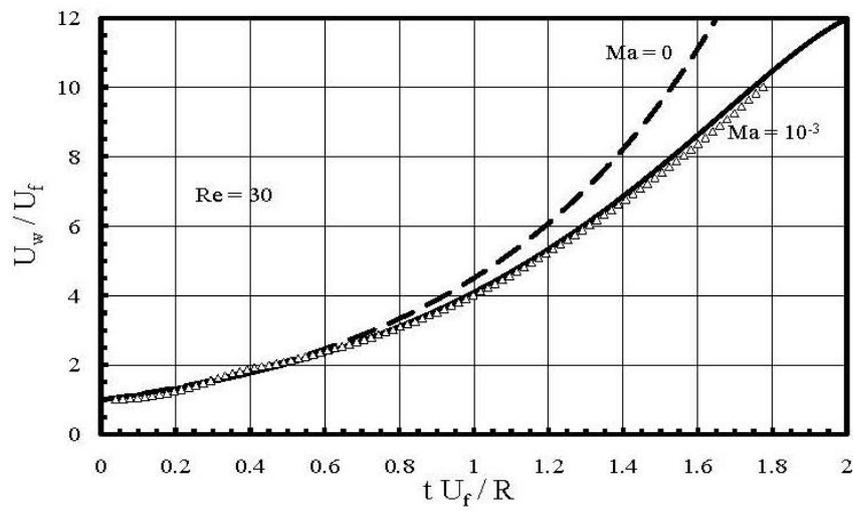

**Figure 2**. Scaled burning rate for $\Theta = 8$, $\gamma = 1.4$, $\mathrm{Re} = 30$ as predicted by the theory for $Ma = 0; 10^{-3}$ (lines) and found in numerical simulations (markers) (Bychkov *et al.* 2010a)

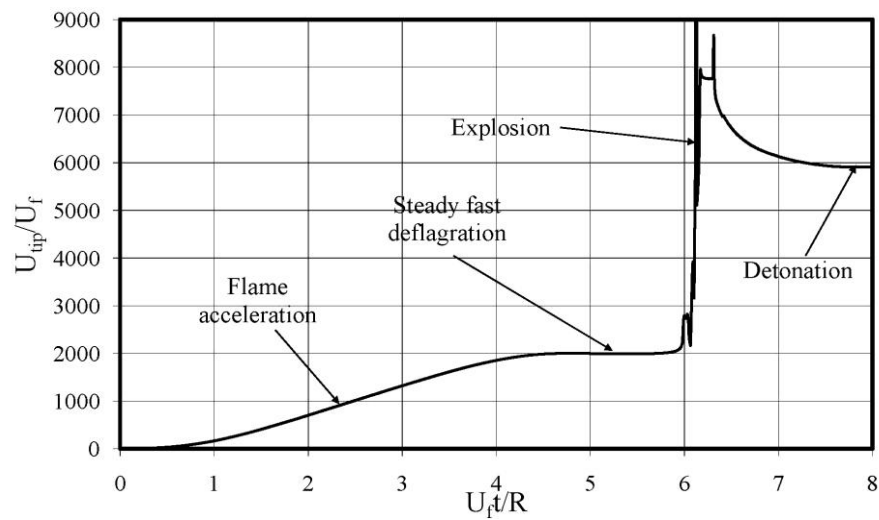

**Figure 3**. Evolution of the flame tip velocity for $\Theta = 8$, $\gamma = 1.4$, $\mathrm{Re} = 6.7$, $Ma = 10^{-3}$ until the full-developed CJ detonation (Valiev *et al.* 2009)

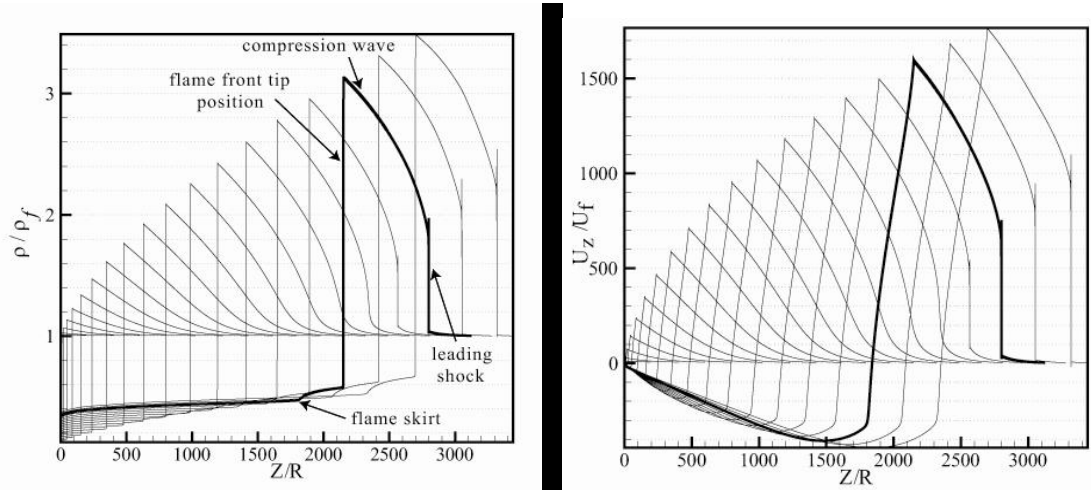

**Figure 4**. Density and velocity profiles along the channel axis for $\mathrm{Re}=6.7$. Time instants are equally spaced in the range of $(0-3.8)U_f t/R$. The plot selected by bold is related to $U_f t/R = 3.28$ (Valiev *et al.* 2009)

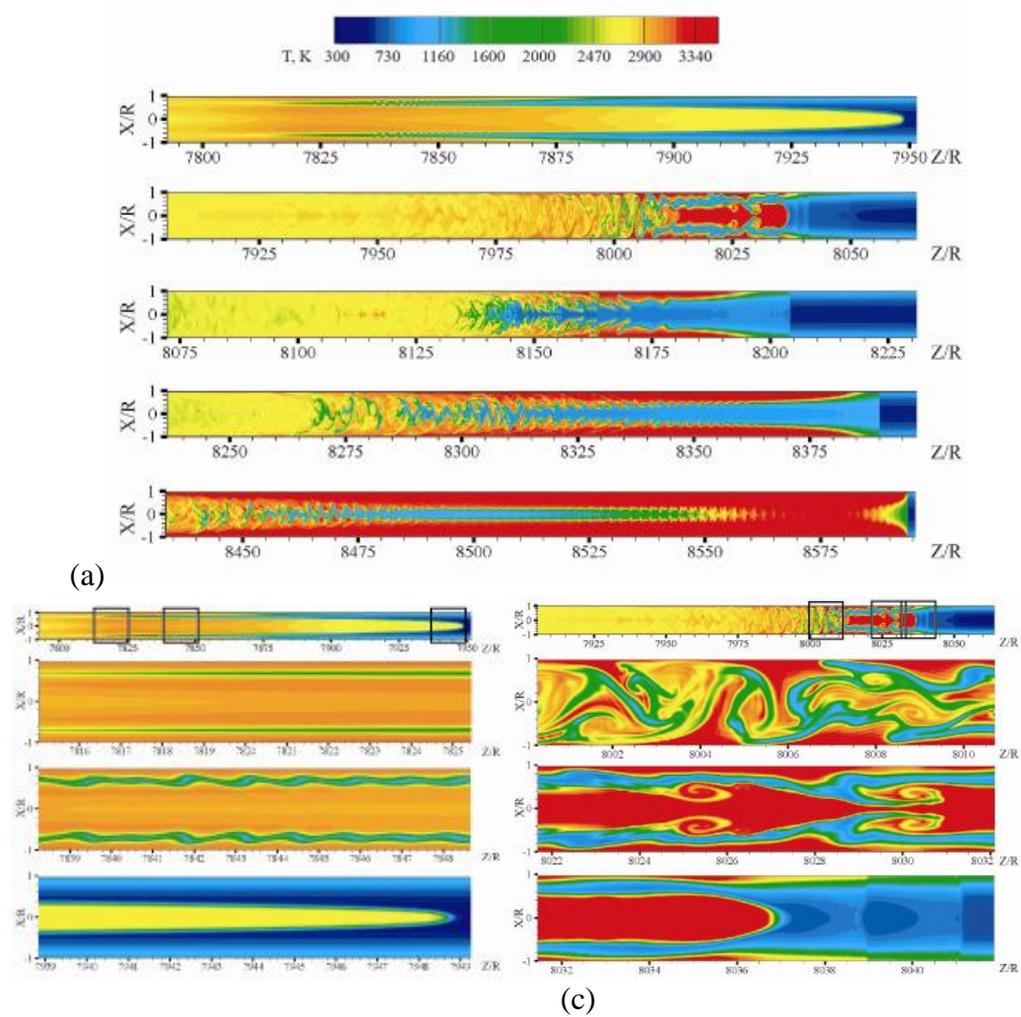

(a)

(b)        (c)

**Figure 5**. Temperature field during DDT for $\mathrm{Re}=10$; (a) Time instants are equally spaced in the range $(7.0-7.16)U_f t/R$; (b) Close-up view with original aspect ratio on time instant $7.0\,U_f t/R$; (c) Close-up view on time instant $7.04\,U_f t/R$ (Valiev *et al.* 2009)

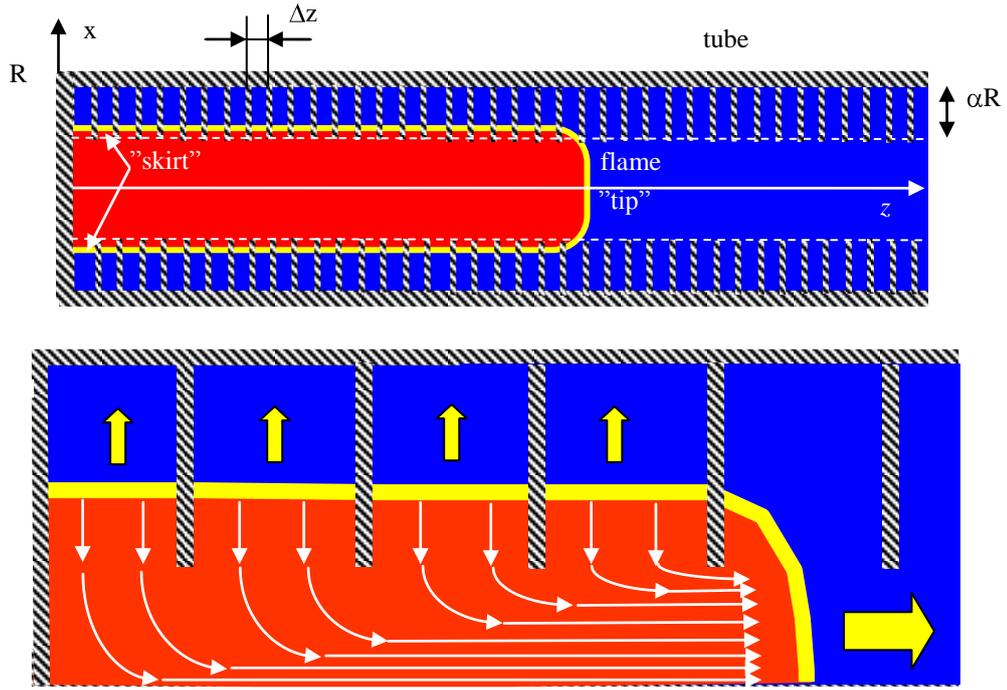

**Figure 6**. Schematic of the physical mechanism of flame acceleration in tubes with obstacles

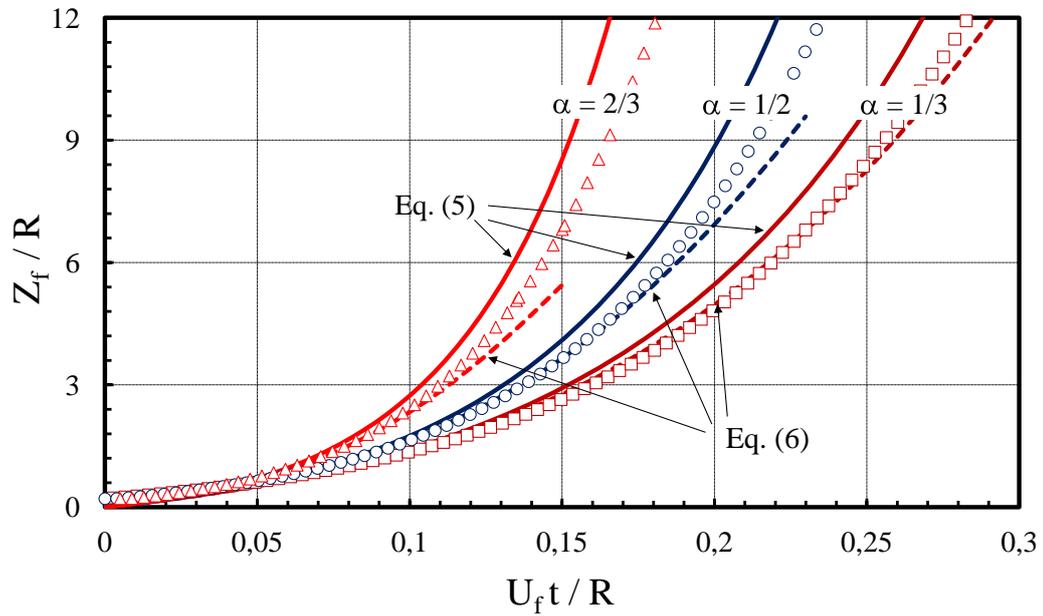

**Figure 7**. Flame tip position for $\Theta = 8$, $\gamma = 1.4$ as predicted by the theory for $Ma = 0; 10^{-3}$ and different values of the blockage ratio $\alpha = 1/3; 1/2; 2/3$ (lines), and found in the numerical simulations (markers) (Byckov *et al.* 2010b)

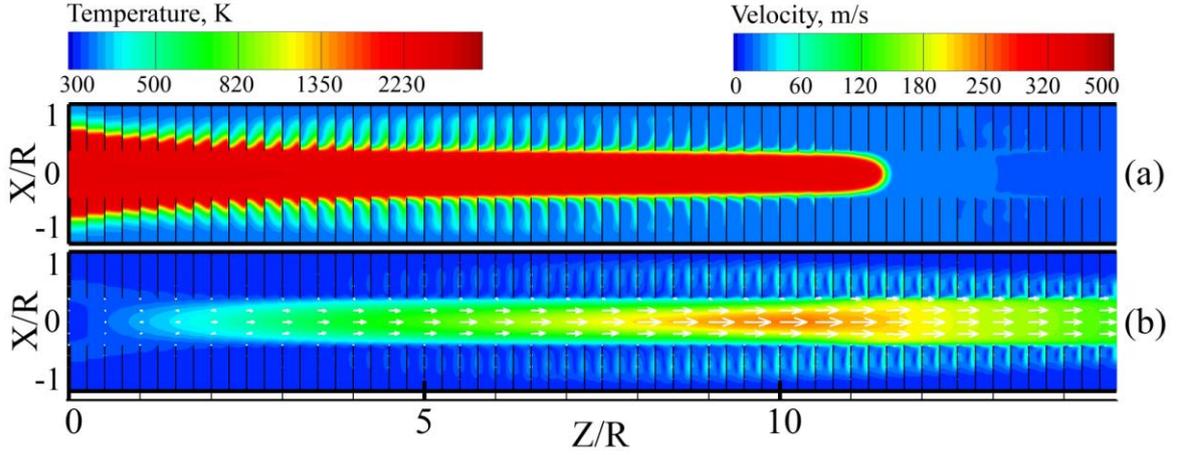

**Figure 8**. Snapshots of temperature (a) and velocity (b) of burning in channels with obstacles for $\Theta = 8$, $Ma = 10^{-3}$, $\alpha = 2/3$, $\Delta z / R = 1/4$ (Bychkov *et al.* 2008)

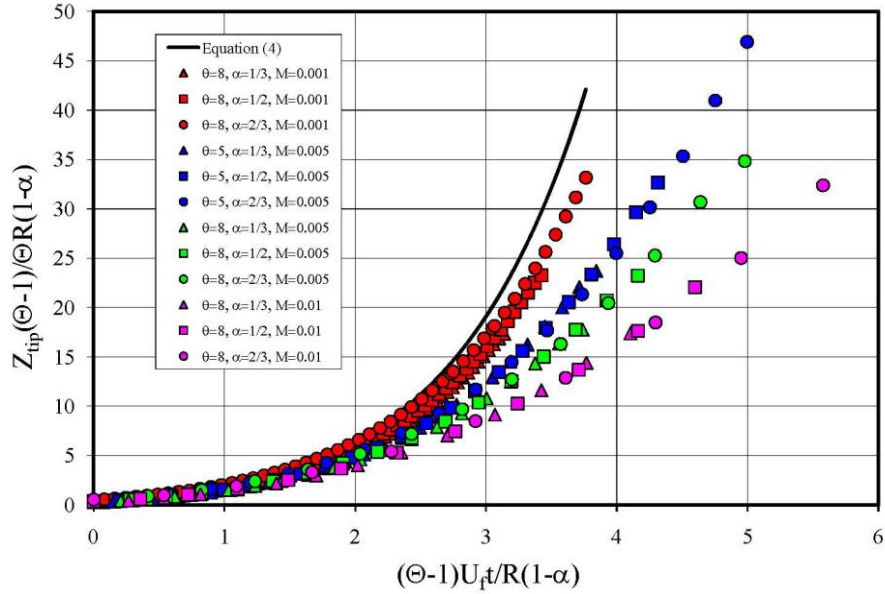

**Figure 9**. Flame tip position as predicted by the theory for $Ma = 0$ and obtained in the simulations for different values of the blockage ratio $\alpha = 1/3; 1/2; 2/3$, $\Theta = 5; 8$ and $Ma = 10^{-3}; 5 \cdot 10^{-3}; 10^{-2}$ (markers) (Bychkov *et al.* 2008)

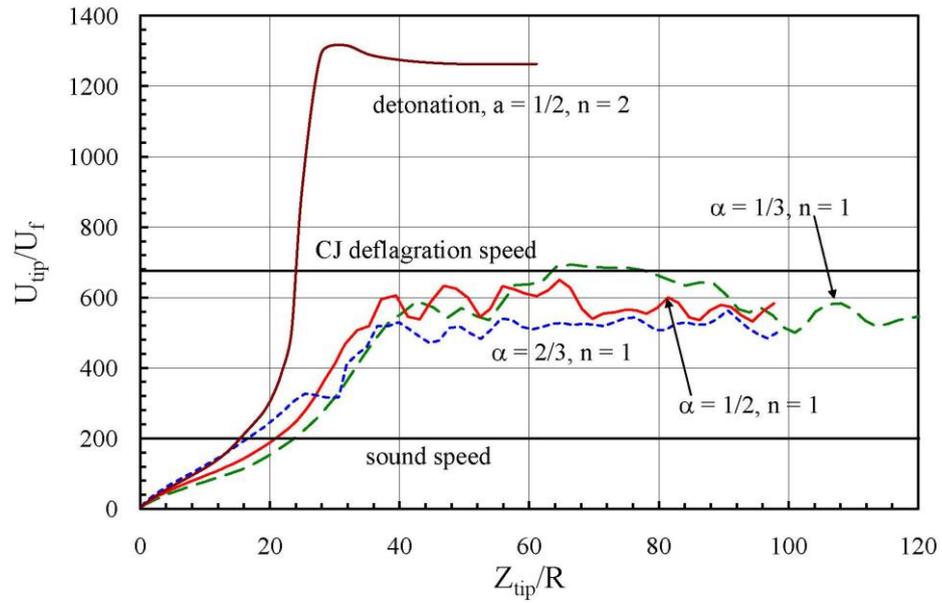

**Figure 10**. Time dependence of the flame tip velocity for $\Theta = 8$, $\alpha = 1/3; 1/2; 2/3$, $Ma = 5 \cdot 10^{-3}$ and various reaction orders with respect to density $n = 1; 2$ (Valiev *et al.* 2010)